\documentclass[reprint, floatfix, pra, nofootinbib, superscriptaddress,aps]{revtex4-1}

\usepackage{graphicx}
\usepackage{bm}
\usepackage{physics}
\usepackage{multirow}
\usepackage[table,dvipsnames,svgnames]{xcolor}
\usepackage[linkcolor=cyan]{hyperref}
\usepackage{svg}
\usepackage{amsmath}
\usepackage{amsfonts}
\usepackage{float}
\usepackage{datetime}
\usepackage{tikz}
\usetikzlibrary{shapes.geometric, positioning}

\begin{document}

\title{Emergence of synthetic twist defects in the surface code under local perturbation}

\author{Paul Kairys}
\email{kairyspm@ornl.gov}
\thanks{\\ 
Notice: This manuscript has been authored by UT-Battelle, LLC, under contract DE-AC05-00OR22725 with the US Department of Energy (DOE). The US government retains and the publisher, by accepting the article for publication, acknowledges that the US government retains a nonexclusive, paid-up, irrevocable, worldwide license to publish or reproduce the published form of this manuscript, or allow others to do so, for US government purposes. DOE will provide public access to these results of federally sponsored research in accordance with the DOE Public Access Plan ( https://www.energy.gov/doe-public-access-plan ).}
\affiliation{Quantum Information Science Section, Oak Ridge National Laboratory, Oak Ridge, TN 37830}
\affiliation{{Quantum Science Center}, Oak Ridge National Laboratory, Oak Ridge, TN 37830, USA}

\author{Phillip C. Lotshaw}
\affiliation{Quantum Information Science Section, Oak Ridge National Laboratory, Oak Ridge, TN 37830}
\affiliation{{Quantum Science Center}, Oak Ridge National Laboratory, Oak Ridge, TN 37830, USA}

\date{\today}

\begin{abstract}
   Topologically-ordered quantum states with Abelian excitations can host defects that obey effective non-Abelian statistics, in principle allowing for quantum information processing via defect braiding. These extrinsic defects (or twists) are typically studied as static features of the lattice. However, an alternative proposal considers how an underlying topologically ordered quantum substrate can be locally perturbed to create and manipulate synthetic defects \cite{you_synthetic_2013}. Unfortunately, while largely referenced, elements of this proposal were never systematically studied. Understanding the energy spectrum is particularly important in finite size and finitely perturbed systems, which are crucial for experimental realizations. In this work we announce a significant step in this direction by explicitly constructing, simplifying, and numerically studying the spectral properties of synthetic defects in a model system. First, we introduce two alternative representations of this problem in both spin and Majorana languages. In the former we describe emergent virtual symmetries which constrain and simplify the problem and in the latter we show a direct connection to Kitaev's well-known Majorana chain. We utilize these simplifications to perform numerical calculations to indicate the location of the quantum phase transition driving the emergence of the synthetic defects. We conclude by discussing key steps for future work to more clearly and completely study this phenomena.   
\end{abstract}

\maketitle

\section{Introduction}\label{sec:intro}

Robustly encoding and manipulating quantum information is crucial to developing quantum technologies for real-world applications. The methods to achieve this goal are generically referred to as quantum error correction (QEC) \cite{terhal_quantum_2015}. It is useful to delineate this broad subject into two complimentary and overlapping topics: \textit{active} or \textit{passive} error correction. The first paradigm operates by repeatedly monitoring a quantum system for errors, deciding what errors have occurred, and systematically acting to undo those errors. The latter aims to encode quantum information and processes so the intrinsic dynamics of the quantum system naturally suppress or restrict errors without direct action from the operator.

One of the most studied, and arguably most promising, avenues for passive error correction is the use of topological order to non-locally encode quantum information \cite{lebreuilly_autonomous_2021}. In addition, some non-Abelian topological phenomena can be used to implement error-resilient quantum information processing through the non-trivial braiding statistics of the systems' topological excitations \cite{stanescu_introduction_2016,terhal_quantum_2015,nayak_non-abelian_2008,wen_choreographed_2019}. 

These possibilities have spurred a tremendous search and engineering of real quantum systems which can intrinsically host these topological properties. One of the most promising is the field of quantum materials where a synthesized material may host topologically ordered ground states. For example, quantum spin liquid materials are envisioned to provide a manipulable substrate hosting a topologically ordered phase on which quantum information tasks can be implemented via local processes \cite{zhou_quantum_2017,savary_quantum_2017,nayak_non-abelian_2008}.

Unfortunately, there remains critical knowledge gaps relating the theoretical concepts of topological quantum information processing with realizations in a laboratory setting. One example, which motivates this work, is the ability to induce non-Abelian topological phenomena within an otherwise Abelian topological system \cite{bombin_topological_2010}. This is most studied in the context of twist defects in the $\mathbb{Z}_2$ surface/toric codes \cite{brown_topological_2013,yan_generalized_2024,kesselring_boundaries_2018,you_non-abelian_2019}. Twists are local geometric defects in the lattice structure of the model which modify the global topological order \cite{barkeshli_symmetry_2019}. 
Twists can be understood in the language of Majorana Fermions \cite{chen_topological_2020}, which we will define and explore later. Within the Majorana representation, the edges of the twists each host a single immobile Majorana Fermion, which are non-Abelian anyons \cite{chen_topological_2020}. In the context of quantum materials, the implementation of twists in a quantum material is a hefty challenge as it would require the precise engineering and synthesis of extended lattice defects in the material's crystalline structure.

An alternative route was proposed by You, Jian, and Wen \cite{you_synthetic_2013} in a model topological spin system, the Wen plaquette model. It consists of applying a local perturbation to exert the same effective action of the twist on the topological excitations and in the strong perturbation limit, to freeze the perturbed spins, inducing an effective lattice dislocation and generating a \textit{synthetic} twist with projective non-Abelian statistics \cite{you_projective_2012,barkeshli_symmetry_2019}. 

You et al. showed using perturbation theory that this modified Wen plaquette model hosts the same stabilizing terms in its Hamiltonian as the original twists \cite{you_synthetic_2013}. This formed the foundation for a new discussion of braiding and manipulation of synthetic twist defects. The significance of this paradigm lies in its potential experimental feasibility: applying local perturbations to manipulate and modify a substrate's topological properties is fundamentally distinct from the meticulous engineering of the microscopic structure of quantum materials. It turns the main challenge from one of material design and engineering, to one of control. It also opens the opportunities to explore new materials which cannot be currently engineered at the atomic level.

While this concept has received much attention, the emergence and properties of synthetic twist defects at fixed system sizes, in bounded time, and with finite-strength perturbations have not been studied. This is critically important because changes in topological properties inherently requires traversing quantum phase transitions \cite{chen_local_2010}. The types of phase transitions that will be observed and how best to mitigate dynamical errors encountered when creating synthetic defects in experiments are thus imperative to determine.

This is not just important for initially preparing the topological phase, but also processing information via anyonic braiding. Braiding requires creating and manipulating excitations in the topological substrate \cite{nayak_non-abelian_2008,chen_local_2010}. How quickly and how robustly these processes can be done is dependent on the precise energy structure of the system. Additionally, any realistic implementation requires understanding the interplay between competing length- and time-scales of perturbations. For example, studying the effect of temperature-dependent fluctuations may require the ability to simulate distributions of variable weight perturbations. All together, these compel us to refine and extend our understanding of synthetic defects.

This work lays the foundation for such analysis by explicitly constructing, simplifying, and numerically studying the emergence of synthetic defects. Our work develops a systematic understanding of the value and practicality of synthetic topological defects through realistic computational modeling.

The rest of the manuscript is structured as follows: in Sec.~\ref{sec:analytic_results} we define the model system and discuss two alternative ways to study this problem, providing multiple routes to numerically approach the problem. Next in Sec.~\ref{sec:numerical_results} we demonstrate that we can exactly diagonalize the Hamiltonian of the model system and using scaling analysis we estimate the location and nature of the phase transition between the two topological orders. We then provide numerical results for a new perturbation geometry that has not been studied in literature before. Finally, we conclude in Sec.~\ref{sec:conclusion} and discuss next steps in the study of synthetic twist defects. 

\section{Model and representations}\label{sec:analytic_results}

\subsection{Wen plaquette surface code}

\begin{figure}
    \centering
    \includegraphics[width=\linewidth]{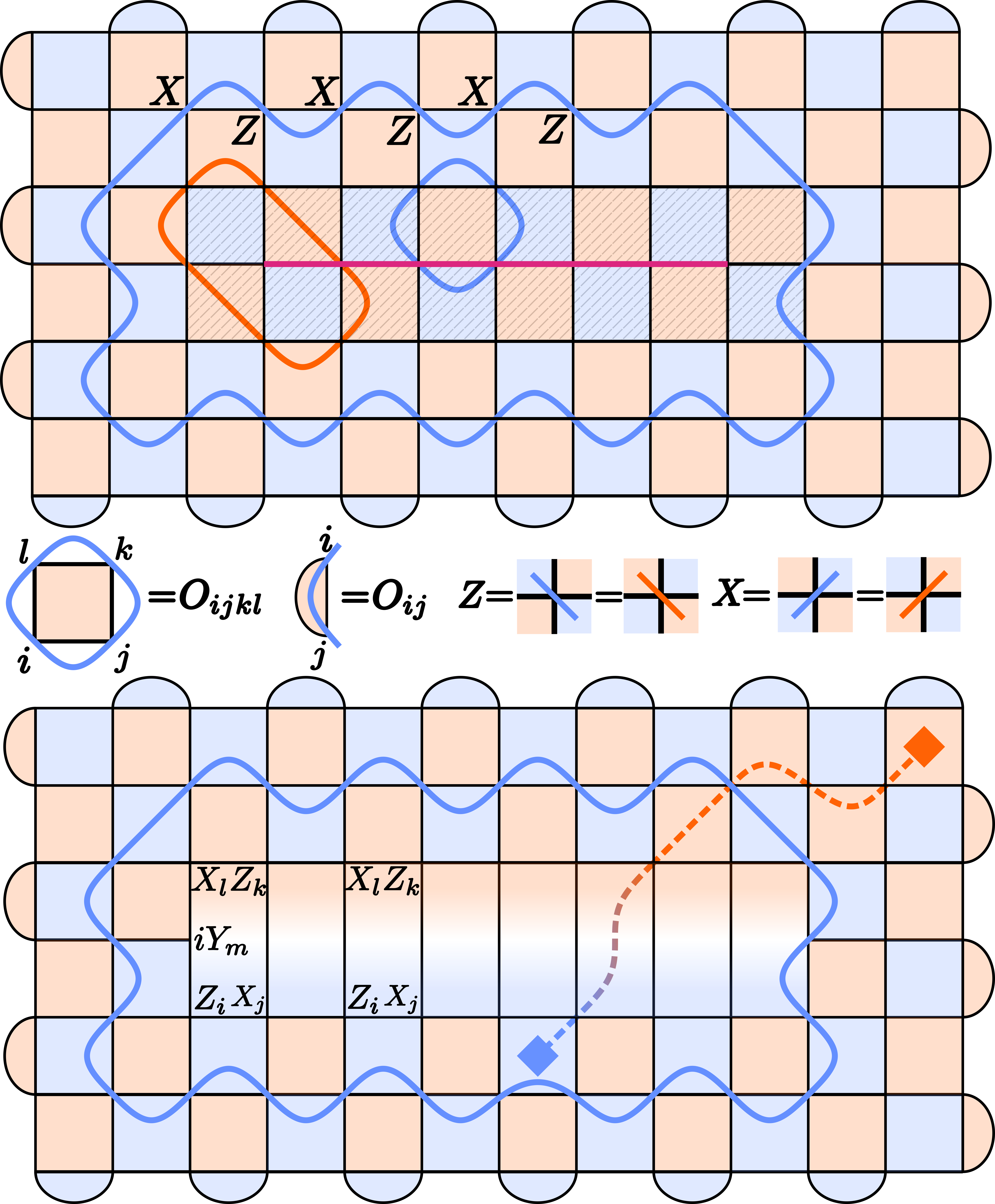}
    \caption{(Upper) The Wen-plaquette surface code showing a set of spins with an applied transverse field. Shown in blue/orange colors are the two sublattices and string operators which are products of $X$ and $Z$ operators on single spins. (Lower) The Wen-plaquette model with a lattice dislocation introducing twists, which are re-defined as four and five-body stabilizer operators on enlarged plaquettes. Anyons (stabilizer excitations) no longer have a globally conserved anyon type (color).}
    \label{fig:lattice_ops_diagram}
\end{figure}

The model system we are interested in is a perturbed version of the Wen plaquette surface code on an $N_x\times N_y$ square spin lattice with a boundary. The set of spins are given by the set $A$ with $|A|=N_x N_y$. We identify two types of operators: 4-spin square plaquette operators $O_{ijkl} =  Z_i X_j Z_k X_l$ and 2-spin boundary operators $O_{ij} = Z_i X_j$, where $X,Y,Z$ are the spin Pauli operators, shown in Fig.\ref{fig:lattice_ops_diagram}. Because the plaquette and boundary operators commute, they are all contained within a single set of stabilizer operators $\Sigma$ where $|\Sigma|=(N_x-1)(N_y-1)+N_x+N_y - 2 = N_xN_y-1$. The unperturbed Hamiltonian is

\begin{equation}\label{eq:wp_ham}
    H = - \sum_{\alpha \in \Sigma}O_\alpha
\end{equation}

The solution to the Wen plaquette model is well understood as an instance of a $\mathbb{Z}_2$ spin liquid, equivalent to the Kitaev's toric code model \cite{you_projective_2012,kitaev_anyons_2006}. This model is also sometimes referred to as the ZXXZ surface code \cite{bonilla_ataides_xzzx_2021}.

The ground states of the unperturbed model are formed analytically by defining the projection on to the eigenstates of $O_\alpha$ with eigenvalue $+1$. Explicitly, let $Q^\pm_\alpha = \frac{1}{2}(I\pm O_\alpha)$ be the projector onto the subspace of eigenvectors of $O_\alpha$ with eigenvalues $\pm 1 $. Then the global ground state $\ket{\psi}$ of the unperturbed Hamiltonian is an eigenvector of $Q^{+} = \prod_\alpha Q^{+}_\alpha$ with eigenvalue $+1$. Thus, the ground state can be interpreted as having satisfied every stabilizer operator in $\Sigma$. 

The degenerate ground states of this model are distinguished by non-contractible loop operators which commute with every operator in $\Sigma$ but are not generated by elements of $\Sigma$, often referred to as logical operators \cite{terhal_quantum_2015}. These logical operators are distinguished from observables in this model called region operators, which are defined by a product of plaquette operators $O_\Gamma = \prod_{\gamma \in \Gamma} O_\gamma$. When the product is taken on either the orange or blue sublattices (shown in Fig.~\ref{fig:lattice_ops_diagram}) these operators define contractible loop operators on the lattice. Contractible and non-contractible loop operators are key to understanding the topology of the underlying model with and without defects/perturbations (See the loops in Fig.~\ref{fig:lattice_ops_diagram}). 

Beyond ground states, the Wen plaquette model hosts excitations that realize Abelian anyonic statistics. An excited state is associated with at least one unsatisfied stabilizer operator. If the region $\Gamma$ contains only orange (blue) plaquettes this corresponds to a blue (orange) loop operator, and the eigenvalue of the operator corresponds to the parity of excitations of orange (blue) stabilizers within the loop.

One can also define twist operators (shown in the lower panel of Fig.~\ref{fig:lattice_ops_diagram}) which bridge the two sublattices \cite{bombin_topological_2010,you_projective_2012}. The boundary created by these operators transmutes a blue excitation (a violation of a blue stabilizer) into an orange excitation (a violation of an orange stabilizer). The edges of the twist boundary interact with the anyonic excitations in a non-Abelian way \cite{bombin_topological_2010}. 

To study virtual twist defects, we define a set of sites $C \subset A$ onto which an external transverse field will be applied -- called the cut (See for example the highlighted pink spins in Fig.~\ref{fig:lattice_ops_diagram}). The perturbed model is then given by

\begin{equation}\label{eq:ham}
    H(\mu,w) = -
    \frac{\mu}{2} \sum_{\alpha \in \Sigma}O_\alpha + w \sum_{i \in C} Y_i
\end{equation}

In this case, the perturbations correspond to the twist boundary from Fig.~\ref{fig:lattice_ops_diagram} and the edges of the boundary are argued to host projective non-Abelian statics analogous to the original twist \cite{you_projective_2012}. 

In the next two subsections we discuss two formally equivalent, but physically distinct, ways to study Eq.~\ref{eq:ham}. 

\subsection{Stabilizer-spin Representation}

One could approach solving this problem by isolating the perturbation and plaquette operators from Eq.~\ref{eq:ham} that do not commute into a single term and diagonalizing the perturbed component directly. This would involve solving a problem involving all the perturbed spins and all their neighboring spins. However, this uses an unnecessarily large Hilbert space, while additional structure can be used to reduce the minimal problem size, as we show below.

We explicitly define the basis which block-diagonalizes this Hamiltonian, then we define a mapping into a simpler virtual spin problem involving fewer spins than the conventional approach and providing new physical interpretations. 

Moving forward, we employ a set-based formalism to emphasize that the analysis is valid for arbitrary geometries of perturbed spins in the cut. We will use standard notation to denote the power set of a set $A$ as $\mathcal{P}(A)$ and the set difference of two sets $A$ and $B$ as $A \setminus B$.

Two sets are particularly important: the set of stabilizers with at least one spin in the cut $C$, $\Theta \subseteq \Sigma$, and the set of stabilizers that do not overlap with the perturbation, $\Phi = \Sigma \setminus \Theta$.

Just as the ground state of the unperturbed model can be identified using a projector where all elements of the stabilizer group are satisfied, we can also utilize projectors onto unique configurations of satisfied/unsatisfied stabilizers to form a convenient basis for our analysis.

We will denote a set of satisfied stabilizers in $\Theta$ as $\theta$ and conversely, unsatisfied stabilizers in $\Theta$ as $\overline{\theta} = \Theta \setminus \theta$. Similarly, satisfied and unsatisfied stabilizers in $\Phi$ as $\phi$ and $\overline{\phi} = \Phi \setminus \phi$, respectively. 

We can thus compactly specify a configuration of satisfied and unsatisfied stabilizers using the pair $(\theta,\phi)$. A projector onto a particular configuration is defined by $Q_{\theta,\phi} = Q_\theta Q_\phi$ where 

\begin{align}
    Q_\theta = \prod_{\alpha \in \theta} Q^+_\alpha \prod_{\alpha \in \overline{\theta}} Q^-_\alpha\\
    Q_\phi = \prod_{\alpha \in \phi} Q^+_\alpha \prod_{\alpha \in \overline{\phi}} Q^-_\alpha
\end{align}

It is straightforward to compute $\Tr(Q_{\theta,\phi}) = 2^{N_xN_y}/2^{|\Sigma|}$ for any configuration $(\theta,\phi)$ and thus each configuration space is spanned by $2$ states that can be labeled by an integer $n = 1,2$. Using the commutative properties of the stabilizers and the orthogonality of the projectors $Q_\alpha$, it can be shown that $\bra{m_{\theta',\phi'}}\ket{n_{\theta,\phi}} = \delta_{nm} \delta_{\theta \theta'}\delta_{\phi \phi'}$. Because there are $2^{|\Sigma|}$ possible configurations $(\theta,\phi)$, the states $ \{ \ket{n_{\theta,\phi}} \}$ form a complete, orthonormal basis for the Hilbert space. 

We now study how the perturbation affects the stabilizer configuration basis. Observe that $\{Y_i,O_{\alpha}\} = 0$ if $i \alpha$. This implies $Y_i Q^+_\alpha = Q^-_\alpha Y_i$ if $i \in \alpha$, meaning that any satisfied stabilizers neighboring a perturbed site $i$ will become unsatisfied (and vice-versa) after application of $Y_i$. Thus $Y_iQ_{\theta,\phi} = Q_{\theta',\phi}Y_i$ and the new configuration $\theta'$ is derived from complete exchange of satisfied and unsatisfied stabilizers around site $i$. This also preserves the parity of satisfied and unsatisfied stabilizers in the region around $i$. 

The fact the perturbation acts locally and exchanges configurations implies that the Hamiltonian in Eq.~\ref{eq:ham} can be separated into blocks with diagonal terms dependent on $\phi, \overline{\phi},$ and $\mu$:

\begin{equation} \label{eq:block_diag_ham}
    H(\mu,w) = \bigoplus_\phi \bigoplus_n \bigg[ H_{\Theta}(\mu,w) + \frac{\mu}{2} (|\overline{\phi}|-|\phi| )\bigg].
\end{equation}

Within each block the basis states depend only on the configuration of stabilizers neighboring the cut, $\Theta$, and the perturbations drive transitions between the basis states associated to each configuration. Therefore, each block Hamiltonian $H_{{\Theta}}(\mu,w)$ has dimension $2^{|\Theta|}$. 

We now introduce a mapping to express the block Hamiltonian as an effective model of interacting virtual spin-$1/2$ particles. This serves two purposes: first it allows us to more concisely define the internal structure of the Hamiltonian within each block with a fewer number of spins. Second the mapping onto virtual spins immediately makes clear the presence of additional symmetries within each block. These virtual symmetries further classify different configurations and permit additional numerical optimizations, enabling even larger modeling efforts. 

\begin{figure}
    \centering
    \includegraphics[width=\linewidth]{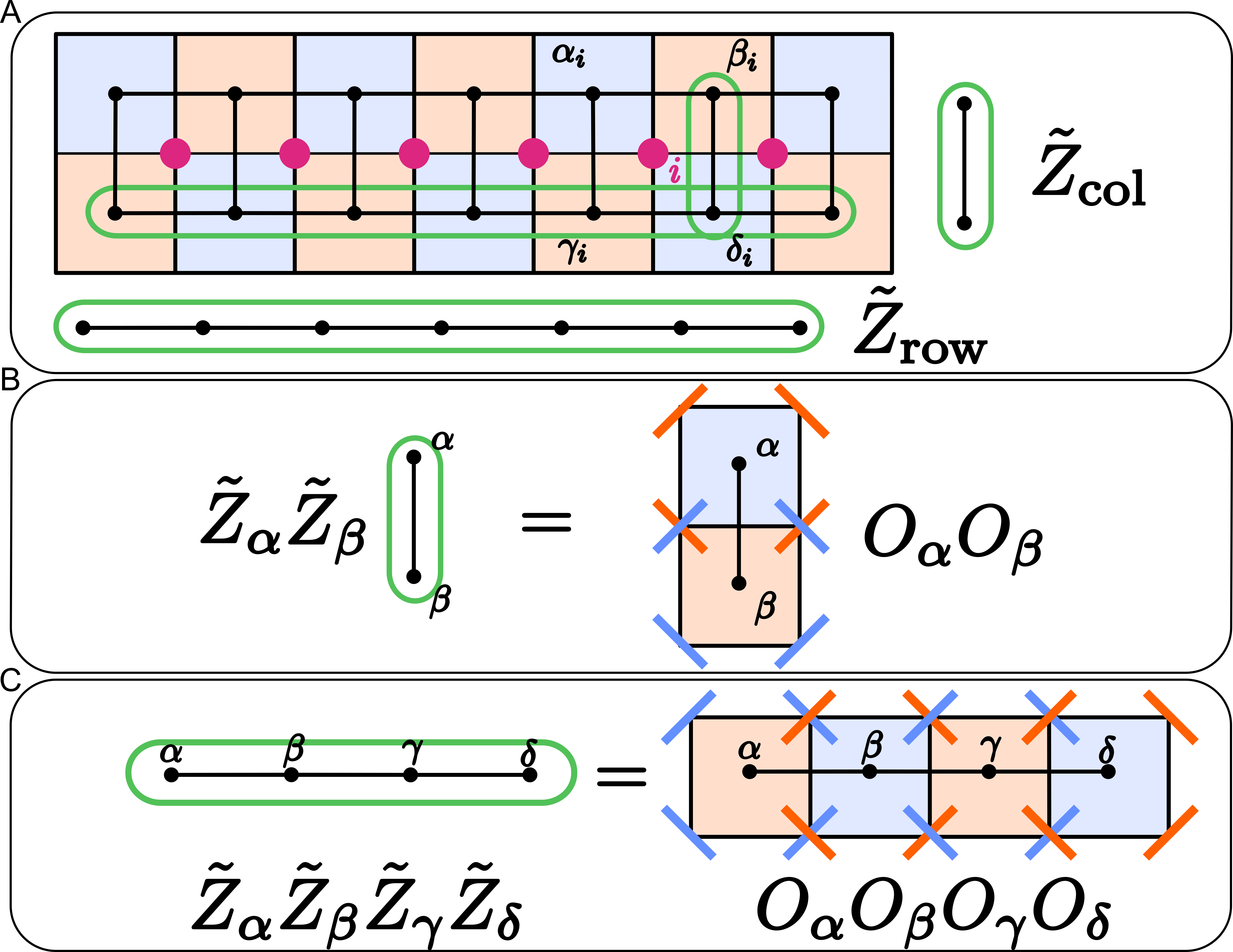}
    \caption{A) The virtual spin problem that emerges from Eq.~\ref{eq:ham} when placing spins (black vertices) on each plaquette face. Virtual row and column symmetries are illustrated as green bubbles around the virtual spins. B) A virtual column symmetry equivalent to the product of the corresponding plaquette operators in the original Wen-plaquette model. C) A virtual row symmetry equivalent to the product of the corresponding plaquette operators.}
    \label{fig:virtual_spin_problem}
\end{figure}

Because the energy of each state is determined uniquely by the configuration of satisfied and unsatisfied plaquettes, we can associate each plaquette with a single variable via a simple map: $s(\alpha)=+1$ if $\alpha \in \theta$ (satisfied) and $s(\alpha) = -1$ otherwise (unsatisfied). Thus, we can obtain a more simple model by associating a virtual spin-$1/2$ degree of freedom for each plaquette (this can be thought of a spin living on each plaquette of the square lattice or in graph-theoretic terms, the dual lattice graph) shown in Fig.~\ref{fig:virtual_spin_problem}A.

We define the mapping first by specifying the local virtual Hilbert space as $\tilde{\mathcal{H}}_\alpha$ and we will define a virtual Pauli-Z operator and its eigenstates as $\tilde{Z}_\alpha \ket{\pm 1} = \pm \ket{\pm 1}$. Then each possible satisfied configuration $\theta$ is associated to a unique basis vector in the composite virtual spin space $\tilde{\ket{\theta}} = \otimes_{\alpha \in \Theta} \tilde{\ket{s(\alpha)}}$ where $s(\alpha)=+1$ if $\alpha \in \theta$ and $s(\alpha) = -1$ otherwise. This mapping is the same for each $n,\phi$, which only serve to label different blocks.

This defines a mapping between basis states in each block: $\ket{n_{\theta,\phi}} \mapsto \tilde{\ket{\theta}}$. To represent the block Hamiltonian in this new representation we can also identify a mapping between operators that respects the algebra of Eq.~\ref{eq:ham}: $O_\alpha \mapsto \tilde{Z}_\alpha$ and $Y_i \mapsto \tilde{X}_{\alpha_i} \tilde{X}_{\beta_i} \tilde{X}_{\gamma_i}\tilde{X}_{\delta_i}$  where $\alpha_i,\beta_i,\gamma_i,\eta_i$ are the plaquettes neighboring spin site $i$ (See Fig.~\ref{fig:virtual_spin_problem}). Using this mapping, each Hamiltonian block can be re-written as

\begin{equation}\label{eq:virtual_ham}
    H_{\Theta}(\mu, w) \mapsto -\frac{\mu}{2}\sum_{\alpha \in \Theta} \tilde{Z}_\alpha + w \sum_{i\in C} \tilde{X}_{\alpha_i} \tilde{X}_{\beta_i} \tilde{X}_{\gamma_i}\tilde{X}_{\delta_i}.
\end{equation}

\noindent We emphasize that this is an explicit representation of the block Hamiltonian defined implicitly in Eq.~\ref{eq:block_diag_ham}.

This explicit representation is directly compatible with standard numerical methods like diagonalization or tensor network techniques. In principal, one could utilize similar methods to simulate Eq.~\ref{eq:ham} directly by integrating the symmetries into the diagonalization or tensor-network algorithms themselves \cite{bridgeman_hand-waving_2017,orus_tensor_2019}, however these additions makes the implementation and optimization of those algorithms more complex. So, the explicit form in Eq.~\ref{eq:virtual_ham} provides a more practical route for numerical study. Aside from numerical implementation, examining the structure of Eq.~\ref{eq:virtual_ham} provides further insights into the physics of the larger model. 

Notably this Hamiltonian obeys virtual string-like symmetries shown in Fig.~\ref{fig:virtual_spin_problem}. This can be used to define a generating set of symmetries that simplify the virtual Hamiltonian further, allowing us to reorder the Hamiltonian into even smaller blocks. The virtual column (Fig.~\ref{fig:virtual_spin_problem}B) and row (Fig.~\ref{fig:virtual_spin_problem}C) symmetry operators described above are equivalent to region operators where the regions are the columns and rows of the plaquettes neighboring the cut. In this context, a state $\ket{\psi}$ violates a virtual symmetry $\tilde{A}$ if $\tilde{A}\ket{\psi} = -\ket{\psi}$ and satisfies the symmetry if $\tilde{A}\ket{\psi} = \ket{\psi}$. 

For a single line of $|C|$ perturbed spins the dual lattice has $2|C|+2$ virtual sites (shown in Fig.~\ref{fig:virtual_spin_problem}). We can define a product operator $\tilde{Z}_\text{row}$ for the top row of virtual spins and $|C|+1$ $\tilde{Z}_\text{col}$ operators for each column of virtual spins. Projecting onto the eigenstates of these operators reduces the size of each block to dimension $2^{|C|}$ for a linear cut.

It is important to highlight the connection of the virtual spin symmetries to the twist operators in Fig.~\ref{fig:lattice_ops_diagram}. The virtual symmetries correspond to products of plaquette operators in the original Wen-plaquette model. The spins between adjacent plaquettes are acted on by a product of single spin operators: $ZX \propto XZ \propto Y$. In the infinitely-strong perturbation limit of Eq.~\ref{eq:ham}, in the ground state the cut spins align with the $Y$ magnetic field in an approximate product state, so they can be traced out and the resulting operators realize the original    double-plaquette and twist stabilizer operators from the lower part of Fig.~\ref{fig:lattice_ops_diagram}. 

\begin{equation}
\vcenter{\hbox{\includegraphics[width=0.85\columnwidth]{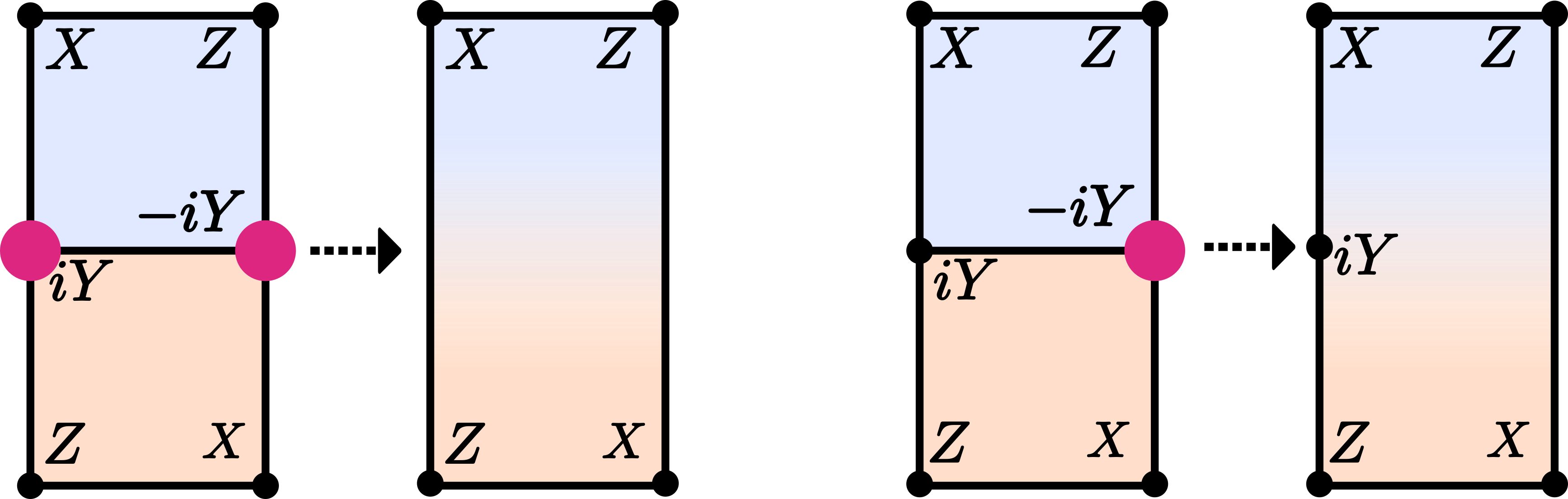}}}
\end{equation}

So, we conclude that the virtual column symmetries correspond precisely to the twist stabilizer operators. 

In contrast, the virtual row symmetry is not contained within the set of stabilizer operators. Instead, in the large perturbation limit, it corresponds to a string operator that commutes with all the stabilizer operators. Thus it corresponds to a new logical operator which distinguishes the emergent ground states.

\begin{equation}
\vcenter{\hbox{\includegraphics[width=0.85\columnwidth]{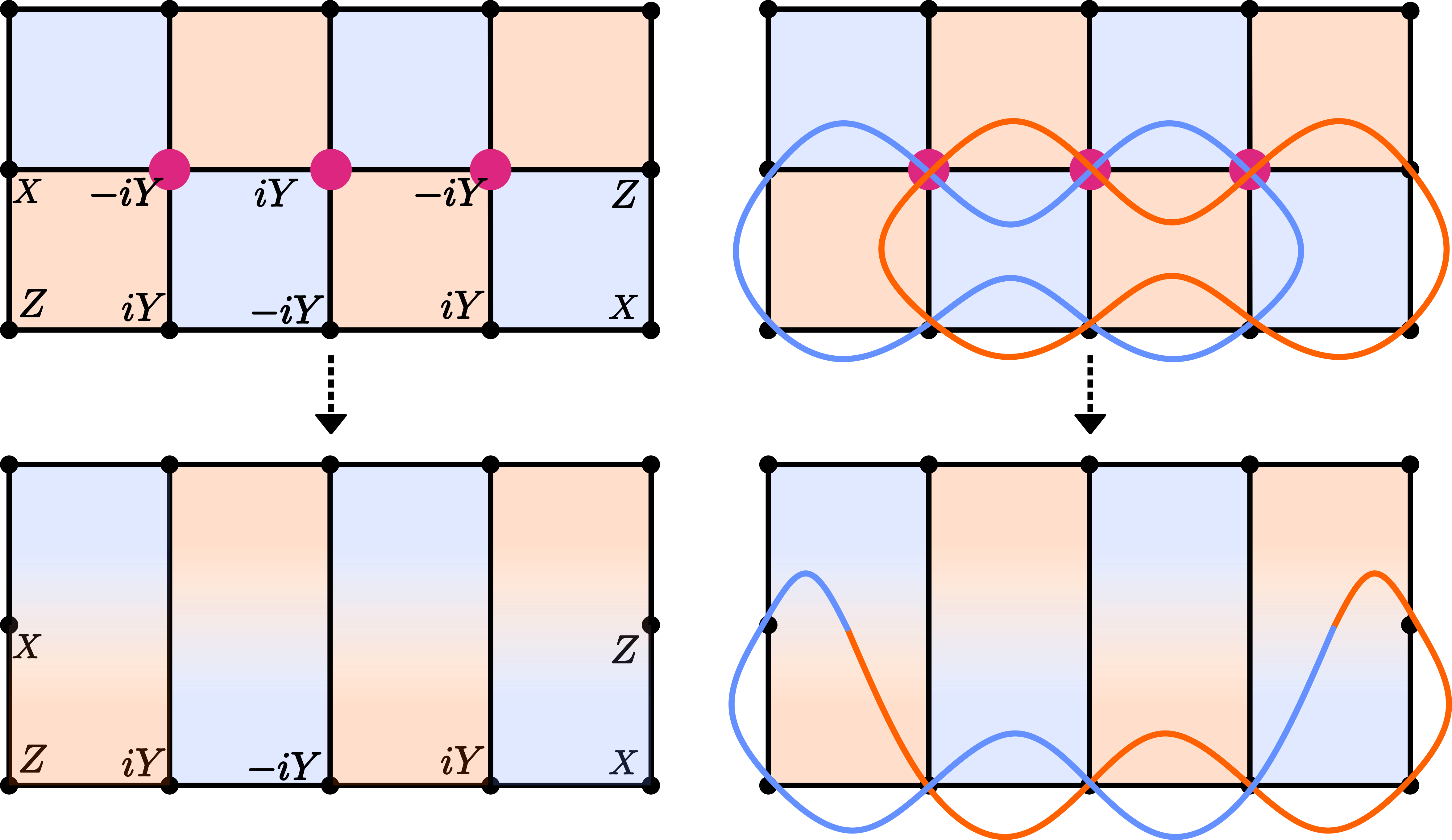}}}
\end{equation}

On the left shows the emergence of the logical operator and the right is a graphical interpretation of the logical operator viewed as a product of two loops with the smallest length that encloses the twist.

It is well known that changing the topological properties of a phase of matter requires traversing a phase transition and a common order parameter used to study phase transitions in spin systems is magnetization. For the virtual Hamiltonian above we can similarly define a local virtual magnetization $M = \sum_{i} \tilde{Z}_i$ as an effective order parameter. This magnetization is equivalent to the expectation value of the local plaquette operators in the perturbed region. A phase with null virtual magnetization would correspond to the plaquette operators no longer approximately stabilizing the ground states of the perturbed Hamiltonian. We will show in later sections that $M$ does a good job of describing the transition to this phase as the field strength increases.

In the next section we discuss an alternative simplification of the model using the formalism of Majorana Fermions. This provides an alternative route for numerical studies and unifies the understanding of defects in the Wen-plaquette model by associating the twists with pairs of isolated non-Abelian Majorana Fermions. 

\subsection{Majorana Fermion Representation}

An alternative route to studying the Wen plaquette model is through a representation of each spin as a set of Majorana Fermions in an enlarged Hilbert space \cite{fu_three_2018}. We utilize the graphical notation introduced in Ref.~\cite{chen_topological_2020} to index and visualize the mapping. 

For each spin $i$ we introduce four-dimensional space with four Majorana Fermions operators $\gamma_i^0,\gamma_i^1,\gamma_i^2,\gamma_i^3$. The Majorana operators are Hermitian $(\gamma_i^a)^\dagger=\gamma_i^a$, involutory $(\gamma_i^a)^2 = 1$, and obey the anti-commutation relations $\{\gamma_i^a,\gamma_j^b\}=2\delta_{ij}\delta^{ab}$. Under a projection operator 

\begin{equation}
    W_i = \frac{I+\gamma_i^0\gamma_i^1\gamma_i^2\gamma_i^3}{2}
\end{equation}

\noindent specific two-Majorana operators can represent the Pauli operator algebra $\sigma^a \sigma^b =\delta_{ab}+i\varepsilon_{abc} \sigma^c$, and can be drawn diagrammatically as:

\begin{equation}
\vcenter{\hbox{\includegraphics{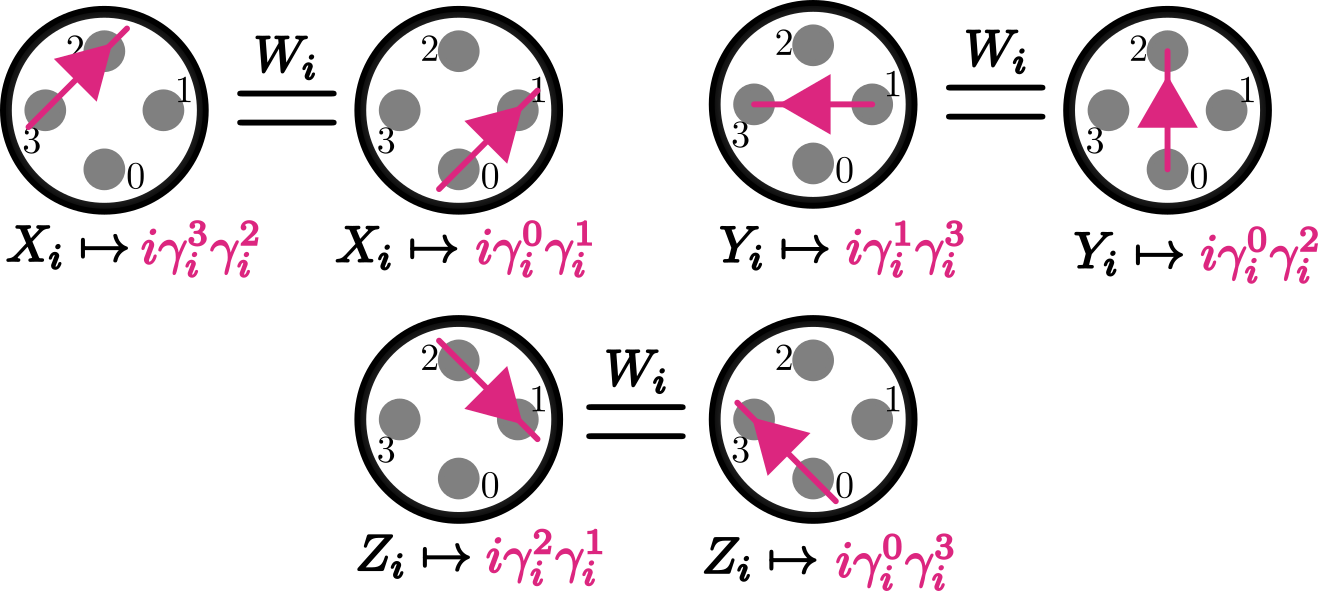}}}.
\end{equation}

\noindent where the arrows indicate the ordering of the Majorana operators, and the $W_i$ over the equality implies equivalence under projection (from here on we omit $W_i$ over diagrammatic equalities to reduce clutter). Importantly, we have the property that $[W_i,\gamma_j^a] = 0$ for $i\neq j$, meaning we can choose either representation for a site and algebraically manipulate the model in the Fermionic space prior to projection onto the spin space. 

Using these definitions, the plaquette operators and boundary operators can be defined as 

\begin{equation}
\vcenter{\hbox{\includegraphics[width=0.85\columnwidth]{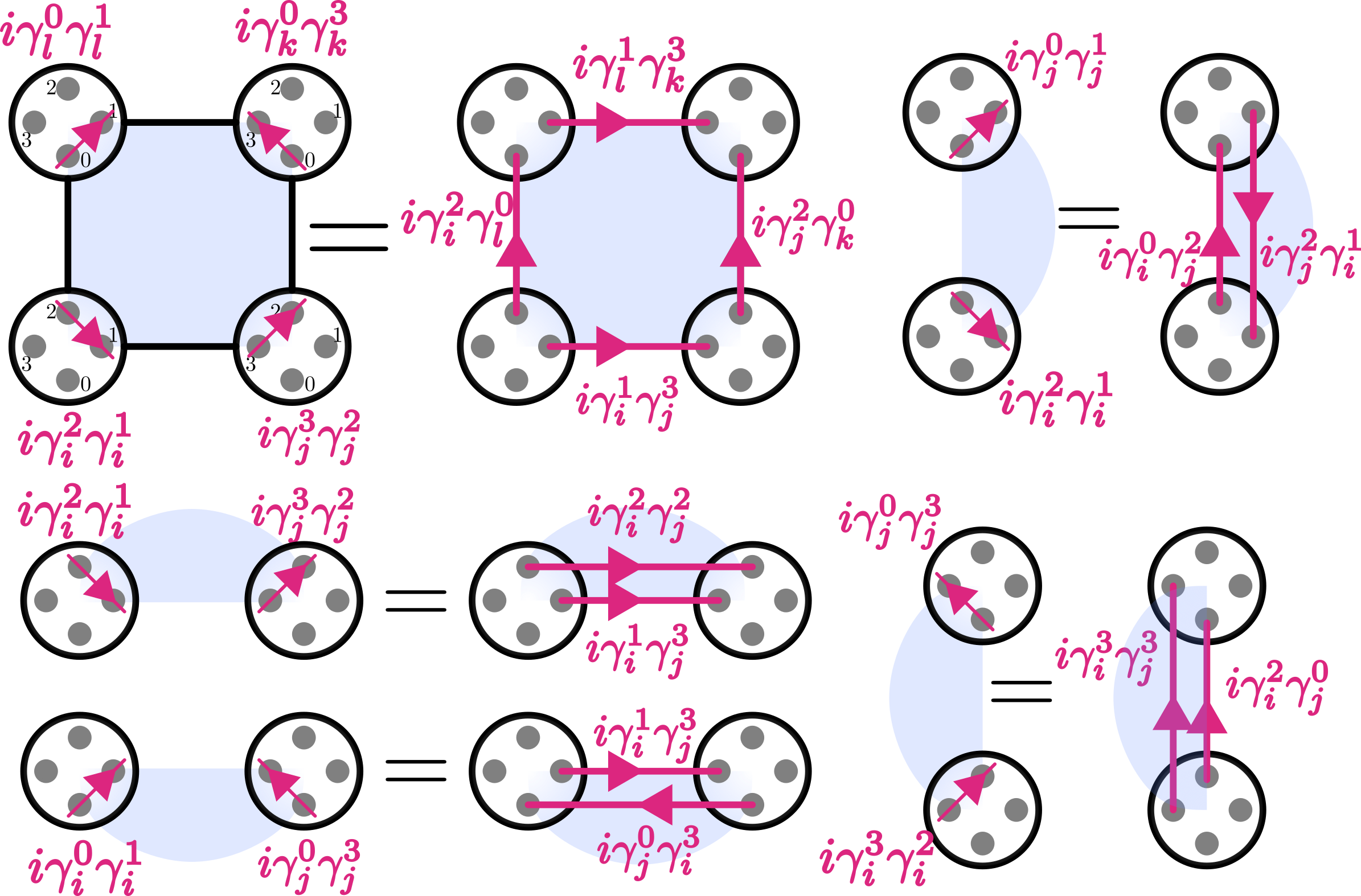}}}
\end{equation}

\noindent where new two-site link operators $\tau_{ij}$ are specified by a product of two Majorana operators from each site and are shown above. Thus, the Hamiltonian (Eq.~\ref{eq:ham}) can be written as 


\begin{align}
    H(\mu,w) = -\frac{\mu}{2} \sum_{\alpha \in \Sigma} \prod_{(ij) \in \alpha} \tau_{ij}+w\sum_i i\gamma_i^1\gamma_i^3. 
\end{align}

It is important to see that each link operator is Hermitian $\tau_{ij}=\tau_{ij}^\dagger$, involutory $\tau_{ij}^2 = I$, and commutes with all other link operators $[\tau_{ij},\tau_{kl}] = 0$. This implies that we can similarly form a basis for the eigenstates of the unperturbed Hamiltonian based on projecting onto different eigenstates of configurations of link operators.

The unperturbed ground state of the Wen-plaquette model is given by the mutual eigenvectors of $\tau_{ij}$ where any even number of link operators around a plaquette are satisfied, corresponding to a satisfied plaquette operator. These different possible satisfying configurations define the degeneracy of the ground states and can be described as an underlying gauge theory \cite{chen_topological_2020}. 


We will now consider, the same 1D line of $N$ perturbed sites as explored above. We can project onto a configuration of $\tau_{ij}$ across the system, labeled $T$, which is analogous to our construction of $\phi,\theta$ previously. Then the perturbation will couple those configurations on which a link operator acts on a perturbed site. This gives rise to a block decomposition of the full Hamiltonian into a link-configuration-dependent Hamiltonian block:

\begin{equation}
    H_T = - \mu \sum_{i=-1}^{N} (-1)^{f(T)}i\gamma_i^1\gamma_{i+1}^3 + w\sum_i^N i\gamma_i^1 \gamma_i^3 + \frac{\mu}{2}(|\overline{\phi}|-|\phi|)
\end{equation}

\noindent where $f(T)$ is the the parity of the configuration of link operators in the plaquettes above and below the edges neighboring perturbed sites and $\phi,\overline{\phi}$ are the number of satisfied and unsatisfied plaquettes outside of the cut, as before. For all of the ground states in the unperturbed model, $f(T)$ is always even. In the space of states that are coupled to these ground states by the perturbation, the Hamiltonian reduces to the well-studied Kitaev chain in a Majorana Fermion representation \cite{kitaev2001unpaired}: 

\begin{equation}\label{eq:majorana_chain}
    H_T =  - \mu \sum_{i=-1}^{N} i\gamma_i^1\gamma_{i+1}^3 + w\sum_i^N i\gamma_i^1 \gamma_i^3.
\end{equation}

\begin{figure*}
    \centering
    \includegraphics[width=\linewidth]{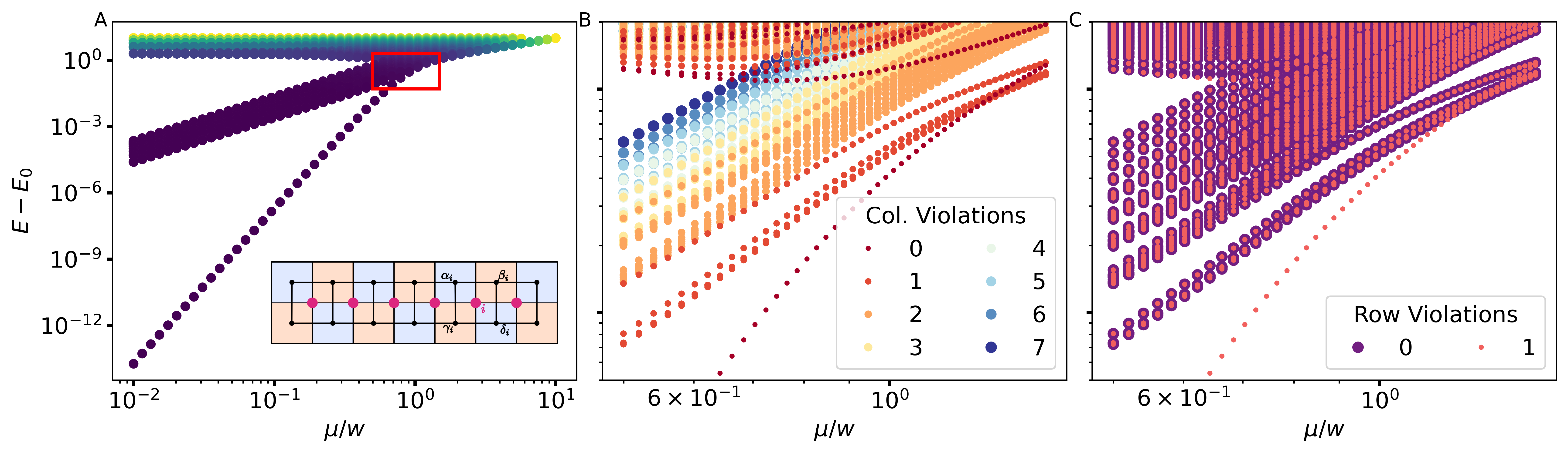}
    \caption{A) The normalized eigenvalues of Eq.~\ref{eq:virtual_ham} for a single line cut of size $|C|=6$ within the $|\overline{\phi}| = 0$, $n=1$ subspace as a function of $\mu / w$ using $w=1$. The color corresponds to the energy on the y-axis. The inset shows the cut size and corresponding virtual spin problem. We clearly observe the emergence of a new ground state in the small $\mu$ limit with a gap that closes exponentially quickly with decreasing $\mu$. We also observe a variety of very low-energy eigenstates. These eigenstates can be partially characterized based on which virtual symmetries are violated. B) A subset of the eigenstates in the red window from $A$ with each state colored by the number of virtual column symmetries that have been violated. C) A subset of the eigenstates in the red window from $A$ with each state colored by the number of virtual row symmetries that have ben violated. We note that the true ground state corresponds to satisfaction of all virtual symmetries and is not shown because the figure uses a logarithmic scale for the $y$-axis and the ground state is normalized to zero energy.}
    \label{fig:spectra}
\end{figure*}

This provides an intuitive explanation for the role of the transverse field perturbation as the potential of pairing of Majorana Fermions on a single spin site, and the plaquette terms from Eq.~\ref{eq:ham} can be seen as a potential to pair Majorana Fermions on the \textit{edges} between sites. Thus, in the strong limit of the perturbation $w$ this gives rise to unpaired Majorana Fermions on the left and right edges of the cut $C$. It is specifically these isolated Majorana Fermions which give rise to the point-like non-Abelian statistics of surface-code twists. 

In this case of a perturbation in a linear chain geometry the effective block Hamiltonian describes $4|C|+8$ Majorana Fermions each with effective dimension $\sqrt{2}$, yielding a total Hilbert space of $2^{2|C|+2}$. There are $|C|+2$ on-site projectors, $W_i$, needed to properly project onto the subspace corresponding to the spin model, giving an effective block dimension of $2^{|C|}$, the same as above for the virtual spin Hamiltonian including the row and column symmetries.

In summary, our analysis exponentially reduces the complexity of studying the energy spectra and properties of the Hamiltonian defined in Eq.~\ref{eq:ham}, within a particular sector of defined excitation (anyon) configuration. In the spin-stabilizer representation these blocks are defined by $(\theta,\phi)$ and in the Majorana representation, $T$. This allows for the study of statics and dynamics for all coupling strengths because the eigenstates and eigenenergies can be determined directly by diagonalization.

\section{Twist Emergence in Simulations}\label{sec:numerical_results}

Thus far we have identified two distinct representations in which to study the effect of a localized transverse field perturbation within the Wen-plaquette surface code model. The first, introduces an effective spin-$1/2$ model with virtual symmetries and the second is a Majorana Fermion representation. In principal, one can utilize either formalism to computationally study the physics of emergent twist defects. For a linear cut geometry, the effective Kitaev chain model can be solved to very large scales via a Nambu formalism \cite{kitaev2001unpaired}. However, the problem becomes more challenging if the cut is not a simple linear geometry, requiring more advanced techniques to simulate large free-Fermion models \cite{schuch_matrix_2019}.

In all of the numerical results that follow we utilize the spin representation for numerical implementations. We first study a finite linear cut geometry then, to demonstrate the flexibility of the spin-based formalism, we study an example of a 2D cut geometry. We utilize exact diagonalization and implement the computations on a MacBook Air with an Apple M2 chip and 16 GB of memory using the Python programming language and open-source numerical libraries.

\subsection{Linear Cut Geometry}

We begin by first studying the low-energy spectrum as a function of $\mu/w$. Using our intuition from the Majorana-Fermion representation (Eq.~\ref{eq:majorana_chain}), we anticipate one phase with Majorana Fermions localized on the edges ($\mu \gg w$) and another phase ($w \gg \mu$) with Majorana Fermions paired along the sites of the cut. We also expect the system to undergo a phase transition at $\mu/w = 1$ in the limit of a large number of sites being cut \cite{kitaev2001unpaired}. 

We report in Fig.~\ref{fig:spectra} the energy gaps for a linear cut of size $|C|=6$ as a function of $\mu/w$ for the $n=1$ subspace. First, we observe that the energy gap between the ground state and first excited state decreases exponentially quickly in the limit $w \gg \mu$, yielding two new ground states (one for each $n=1$ and $n=2$) in the limit of a large perturbation. However, the new ground state of the system only has a vanishing gap in this limit. It is also interesting to note a band of excited states above the new ground state which also have rapidly decreasing energy gaps. 

We noted originally that the virtual symmetries discussed in Sec.~\ref{sec:analytic_results} allow us to reduce the computational cost of studying this system. And they serve an additional purpose: the number of violated virtual symmetries characterizes the low-energy excited states. This can be seen by noting the different shades and colors in Fig.~\ref{fig:spectra} which are related to the number of virtual symmetry operators of either the row or column types which are violated.

The distinction between the column and row virtual symmetries is critical because the new, emerging ground state violates the virtual row symmetry, whereas original ground state does not. As discussed in Sec.~\ref{sec:analytic_results} the column symmetries correspond to new stabilizers and the row symmetry corresponds to a new logical operator. 

We noted in Fig.~\ref{fig:spectra} that there are multiple eigenvectors with eigenvalues that decrease exponentially with decreasing $\mu / w$. While one does converge at a significantly faster rate it is not obvious from this $|C|=6$ finite-size case that this new state alone will characterize the new emergent ground state. Thanks to the connection to the Majorana representation, we anticipate that two new ground states should emerge and the energy gap between the ground state and the excited state should decrease exponentially with increasing system size \cite{kitaev2001unpaired}.

\begin{figure}
    \centering
    \includegraphics[width=0.75\linewidth]{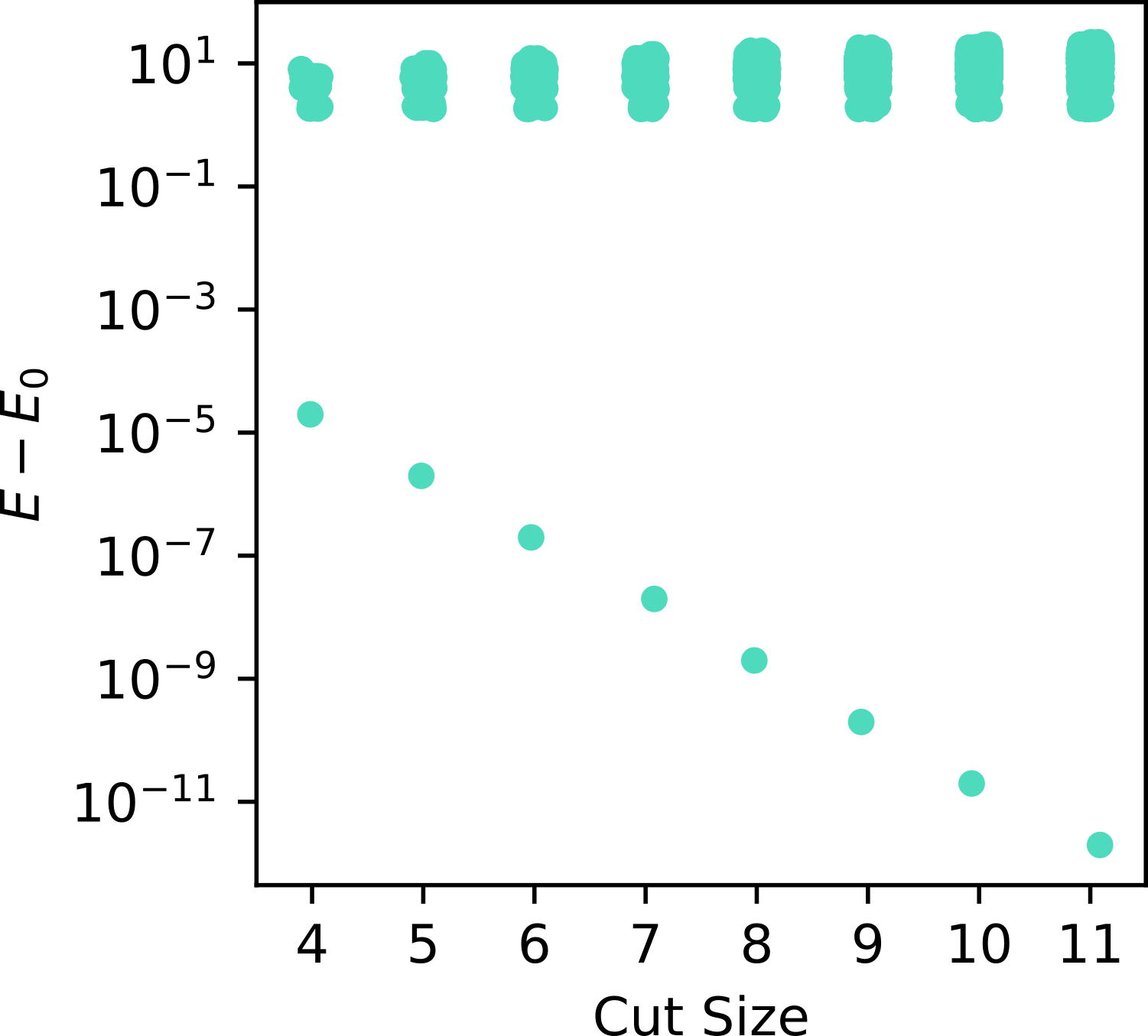}
    \caption{The normalized eigenvalues of Eq.~\ref{eq:virtual_ham} within the $|\overline{\phi}| = 0$, $n=1$ subspace as a function of increasing cut size at a value of $\mu / w = 0.1$. We see that the energy gap between the ground state and the first excited state decreases exponentially quickly as a function of increasing chain length, even at a finite $\mu/w$.}
    \label{fig:gap_fss}
\end{figure}

\begin{figure}
    \centering
    \includegraphics[width=0.75\linewidth]{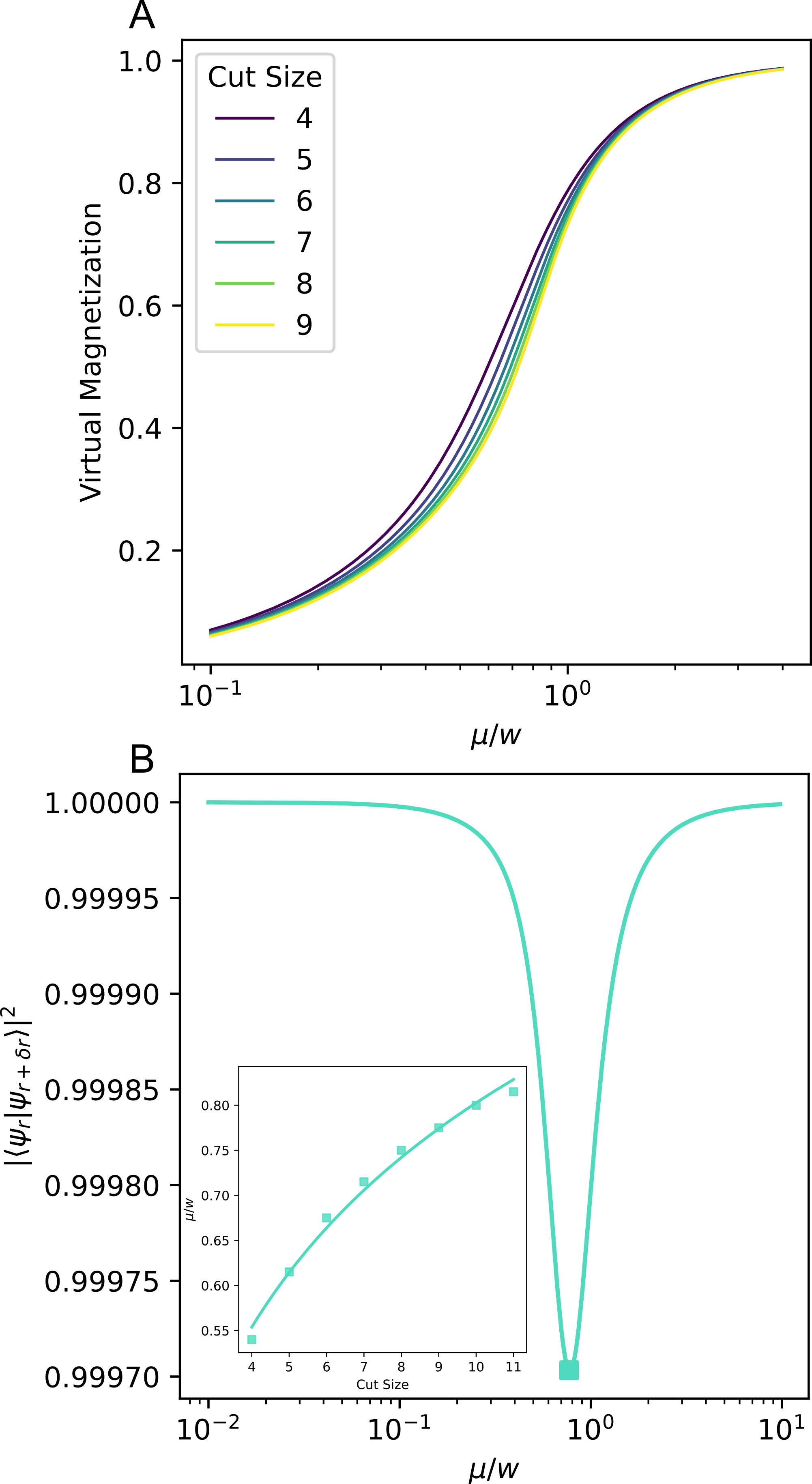}
    \caption{A) The virtual magnetization for the ground state of Eq~\ref{eq:virtual_ham} with increasing cut size. B) The fidelity change in the ground state as a function of $\mu/w$, we observe a minimum and scaling analysis shows the location of this minimum shifts towards the expected phase transition point of $\mu/w=1$.}
    \label{fig:magnetization_fidelity_scaling}
\end{figure}

In Fig.~\ref{fig:gap_fss} we analyze the magnitude of the gaps as a function of cut size $|C|$ at a finite value of $\mu / w = 0.1$. We observe the same exponentially vanishing gap with increasing chain length that one would anticipate from a Kitaev chain \cite{kitaev2001unpaired}. This provides strong numerical evidence that this state is the unique new ground state in the thermodynamic limit and that the new phase remains gapped.

So far we have provided analytic and numeric evidence that new ground states appears when $w \gg \mu$ and that this ground state becomes degenerate exponentially quickly in both increasing $|C|$ and decreasing $\mu / w$. Additionally, it can be distinguished from the original by a virtual string-like symmetry. However from our analysis so far, it is not clear precisely where the transition between these two phases occurs. 

Based on the reduction to a Kitaev chain model shown earlier we would expect to find a phase transition at $\mu/w = 1$ \cite{kitaev2001unpaired,verresen_one-dimensional_2017}. This is observed approximately in Fig.~\ref{fig:spectra} in the location of the excited state level crossings. In order to get a better understanding of the location of the phase transition we examine two quantities. First, we consider the virtual magnetization $\langle M \rangle  = \langle \sum_\alpha \tilde{Z}_\alpha \rangle$ as an effective order parameter and study the the smoothness of this quantity as we vary $\mu / w$. 

In Fig.~\ref{fig:magnetization_fidelity_scaling} A) we show the virtual magnetization as a function of $\mu / w$ for increasing cut sizes. We notice immediately that the transition is quite smooth. And as one increases the size of the cut $|C|$ the the transition in magnetization does not get appreciably sharper. This provides indication of a continuous phase transition in the thermodynamic limit, which is exactly what one expects to find for the standard Kitaev chain \cite{verresen_efficiently_2021}.

Finally, rather than an order parameter we consider another diagnostic of phase transitions, the fidelity \cite{you_fidelity_2007,gu_fidelity_2010,garnerone_fidelity_2009,tang_unveiling_2021}. The state fidelity is an operational measure of how distinguishable two quantum states are from one another. Fidelity has been used as an information-theoretic metric to identify and characterize quantum phase transitions agnostic to an underlying order parameter \cite{zanardi_information-theoretic_2007,garnerone_fidelity_2009}.

We consider a single parameter $r=\mu / w$ which defines the relative energy scale in the Hamiltonian. We consider small changes ($\delta r = 0.001$) in this parameter and compute the fidelity of two nearby ground states $|\langle \psi_r| \psi_{r+\delta r}\rangle|^2$. A minimum in the fidelity change is known to correspond to the location of a phase transition \cite{gu_fidelity_2010}. We show the fidelity change as a function of the parameter $r$ for the ground state of a $|C|=6$ linear system in Fig.~\ref{fig:magnetization_fidelity_scaling}B and we identify a small, but numerically meaningful, dip in the fidelity near the suspected critical point of $r=1$. 

We observe that the minimum in the fidelity decreases with increasing system sizes and (shown in the inset of Fig.~\ref{fig:magnetization_fidelity_scaling}B) we see that the location of the minimum also shifts towards the expected critical point at $r = \mu / w =1$ \cite{kitaev2001unpaired}. While we observe that the location of the minimum approaches the expected critical point, we can not identify how quickly this occurs. Indicating that using fidelity to study the nature of this phase transition may require incorporating states beyond the ground state and analyzing higher-order effects \cite{gu_fidelity_2010,tang_unveiling_2021}.

\subsection{Rectangular Cut Geometry}


\begin{figure*}
    \centering
    \includegraphics[width=\linewidth]{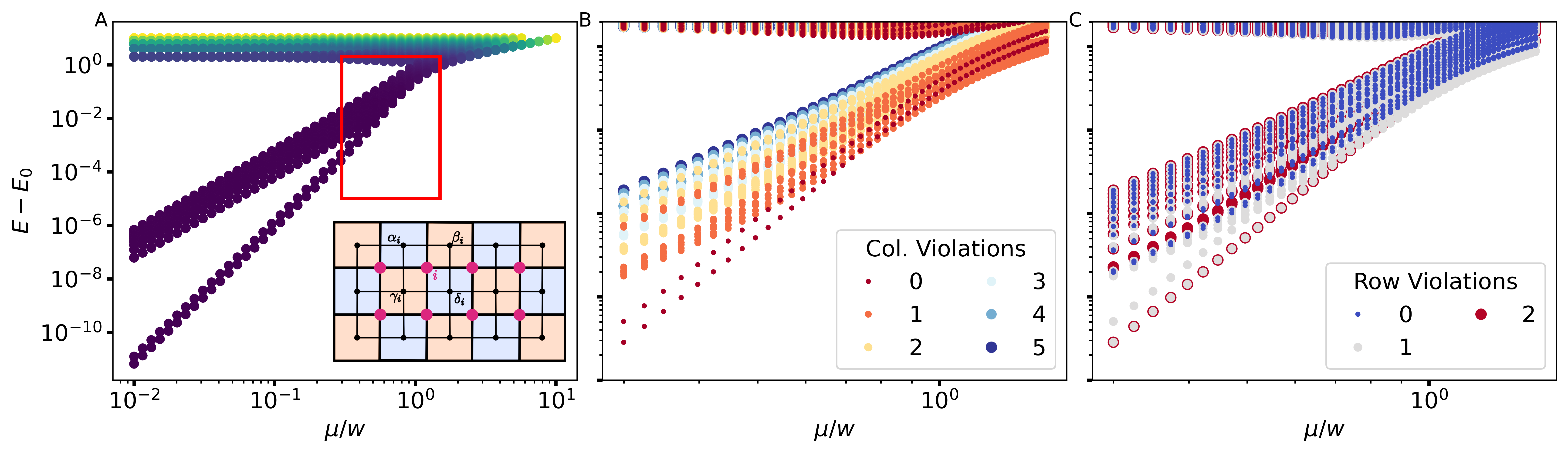}
    \caption{A) The normalized eigenvalues of Eq.~\ref{eq:virtual_ham} for a rectangular cut within the $|\overline{\phi}| = 0$, $n=1$ subspace as a function of $\mu / w$ using $w=1$. The color corresponds to the energy on the y-axis. The inset shows the cut geometry and corresponding virtual spin problem. We observe the emergence of a \textit{2} new ground states in the small $\mu$ limit (the points overlap and can be most easily distinguished in panel C) with a gap that closes exponentially quickly with decreasing $\mu$. We also observe a variety of very low-energy eigenstates, including one very close in energy to the two ground states. These eigenstates can be characterized based on which virtual symmetries are violated. B) A subset of the eigenstates in the red window from $A$ with each state colored by the number of virtual column symmetries that have ben violated. C) A subset of the eigenstates in the red window from $A$ with each state colored by the number of virtual row symmetries that have ben violated. We note that the new ground states corresponds to a either a single or double violation of row symmetries. The original ground state corresponds to zero virtual symmetry violations is not shown because the figure uses a logarithmic scale for the $y$-axis and the ground state is normalized to zero energy.}
    \label{fig:4_2_rect_spectra}
\end{figure*}

Up until now, we have only considered the situation of the perturbation applied to a single linear cut geometry. However, it is interesting to consider other geometries both from a scientific and practical perspective. For the latter case, applying a local magnetic field to a topological substrate will always perturb a finite 2D or 3D volume of spins, not just a 1D line. Understanding how this situation affects the emergence or robustness of synthetic twist defects, or even if something entirely unexpected emerges, is critical to real applications.

To that end, we briefly consider an example of a rectangular cut geometry, which is only minutely more realistic than the initial case, but illustrates the necessity of further study. This more experimentally realistic scenario also emphasizes the utility of the spin-based formalism developed in this work.

We consider on-site perturbations arranged in a rectangular geometry of four perturbed sites in two adjacent rows. This requires 15 plaquettes interacting and therefore the equivalent model requires 15 spins, as shown in the inset of Fig.~\ref{fig:4_2_rect_spectra}A. The virtual Hamiltonian has $5$ column symmetries and $2$ row symmetries.

We show in Fig.~\ref{fig:4_2_rect_spectra} the low-energy eigenvalues of for the ground state configuration for perturbed rectangular cut. In the limit of small $\mu/w$ we observe the emergence of \textit{2} new states with exponentially small gap between the original ground state (these points overlap very closely so are difficult to distinguished in the figure). These states can be distinguished by the number of virtual row symmetry violations. An additional low-energy state is also present, however compared to the lowest two it is not clear if this state is a finite-size artifact. In either case, this is a stark contrast to the linear cut geometry where a total of two new ground states appeared, in this case we observe a total of \textit{4} new ground states (2 for each $n=1,2$ subspace). 

The ground-state degeneracy of the associated topological model corresponds to twice the number of non-contractible loop operators \cite{you_projective_2012}. In our first example of a linear cut geometry, the non-contractible loop operator is effectively given by the virtual row operator, which exchanges blue and orange anyons across the cut. However, in the 2D example, there is an additional row operator. Each row operator allows the anyons to change color and thus an anyon moving across both rows will return to its original color. This interpretation is consistent with our numerical observations and gives us an intuitive topological interpretation of the results.

However, it is now clearly shown that understanding how synthetic twist defects emerge in finite systems and the resulting changes in topological properties are dependent not just on the topology of the underlying lattice, but on the \textit{geometry} of the perturbations as well. Using the techniques developed in this work, it is now feasible to systematically study this phenomena and potential applications.

\section{Conclusion}\label{sec:conclusion}

Engineering and controlling the topological properties of quantum systems is of tremendous importance in the race for robust quantum technologies. One plausible strategy to accomplish this in systems like quantum spin liquids is by engineering synthetic twist defects via local perturbations on an underlying entangled substrate \cite{you_synthetic_2013}. Unfortunately, the emergence and properties of synthetic twist defects have received limited attention, especially in finite-size and finite-perturbation-strength cases. 

To address this, first, we provided a mapping of the perturbed Wen plaquette surface code to two distinct physical systems: a spin-$1/2$ model and a Majorana Fermion model. In the former, we have identified virtual symmetries that characterize the emergent ground states and their low-energy excitations. In the latter, we have explicitly mapped the emergence of synthetic twist defects to the pairing and localization of Majorana Fermions. Finally, in both cases we have shown the computational challenge of studying these twist defects can be exponentially reduced and therefore are amenable to study in realistic scenarios. 

Next we completed numerical calculations to provide insight into the emergence of synthetic twist defects. This includes directly estimating the low-energy spectrum of the Hamiltonian, effective order parameters and symmetries, and conducting scaling analysis to estimate the nature and location of the phase transition leading to synthetic twist defects. We find good agreement between our numerical results utilizing the spin formalism and the expected ground-state properties from our analysis in the Majorana fermion representation.

A number of important extensions to this work have not been explored but are critical to understand the potential and utility of synthetic twist defects. The first potential route is to study the role of cut geometry. 

In this work we have numerically studied two types of cut geometry: a linear change where a single line of sites is perturbed and a rectangular patch where a group of neighboring sites are affected. The first, allowed us to verify the numerical results agreed with the theories and continuum limits expected. Specifically, that the perturbation induces a two new ground states and with exponentially decreasing energy in perturbation strength, and in cut size. However, the second example shows that subtle changes in the geometry of the perturbations can lead to the appearance of even more new ground states. It is clear that understanding the geometry of the perturbations will be critical in assessing the utility of synthetic defects.

One can also now study dynamical properties of emergent twist defects because we have shown the ability to directly compute the eigenvectors and eigenvalues of the system and have even identified virtual symmetries which correspond to conserved constants of motion. In particular, it will be critical to understand how the topological properties of the model can be manipulated/utilized via dynamical processes to store and manipulate quantum information. It will also be critical to understand how the spatial homogeneity of the perturbation will affect these variables. 

We have taken a critical step in the study of synthetic defects by providing explicit constructions and simplifications of the model to the point where numerical analysis is feasible. And we have provided the first numerical study of synthetic defects in finite size and finite-perturbation regimes. Ultimately, this work lays the foundation to systematically investigate the phenomena of synthetic defects in realistic scenarios and ultimately, provide routes to assessing their value for applications in quantum information science. 

\section{Acknowledgements}
The authors gratefully acknowledge Eugene Dumitrescu, Ryan Bennink, and Ammar Ali for invaluable discussions. P.K. and P.C.L were supported by the Quantum Science Center (QSC), a National Quantum Science Initiative of the Department Of Energy (DOE), managed by Oak Ridge National Laboratory (ORNL), for project conception, theory development, simulations, analysis, and writing. 






\bibliographystyle{unsrt}
\bibliography{references}

@article{kitaev2001unpaired,
  title={Unpaired Majorana fermions in quantum wires},
  author={Kitaev, A Yu},
  journal={Physics-uspekhi},
  volume={44},
  number={10S},
  pages={131},
  year={2001},
  publisher={IOP Publishing}
}

@article{you_synthetic_2013,
	title = {Synthetic non-{Abelian} statistics by {Abelian} anyon condensation},
	volume = {87},
	url = {https://link.aps.org/doi/10.1103/PhysRevB.87.045106},
	doi = {10.1103/PhysRevB.87.045106},
	number = {4},
	urldate = {2025-01-03},
	journal = {Physical Review B},
	author = {You, Yi-Zhuang and Jian, Chao-Ming and Wen, Xiao-Gang},
	month = jan,
	year = {2013},
	note = {Publisher: American Physical Society},
	pages = {045106},
}

@article{savary_quantum_2017,
	title = {Quantum spin liquids: a review},
	volume = {80},
	issn = {0034-4885, 1361-6633},
	shorttitle = {Quantum spin liquids},
	url = {https://iopscience.iop.org/article/10.1088/0034-4885/80/1/016502},
	doi = {10.1088/0034-4885/80/1/016502},
	language = {en},
	number = {1},
	urldate = {2021-03-05},
	journal = {Reports on Progress in Physics},
	author = {Savary, Lucile and Balents, Leon},
	month = jan,
	year = {2017},
	pages = {016502},
}

@article{nayak_non-abelian_2008,
	title = {Non-{Abelian} anyons and topological quantum computation},
	volume = {80},
	issn = {0034-6861, 1539-0756},
	url = {https://link.aps.org/doi/10.1103/RevModPhys.80.1083},
	doi = {10.1103/RevModPhys.80.1083},
	language = {en},
	number = {3},
	urldate = {2019-02-08},
	journal = {Reviews of Modern Physics},
	author = {Nayak, Chetan and Simon, Steven H. and Stern, Ady and Freedman, Michael and Das Sarma, Sankar},
	month = sep,
	year = {2008},
	pages = {1083--1159},
}

@article{wen_choreographed_2019,
	title = {Choreographed entanglement dances: {Topological} states of quantum matter},
	volume = {363},
	issn = {0036-8075, 1095-9203},
	shorttitle = {Choreographed entanglement dances},
	url = {http://www.sciencemag.org/lookup/doi/10.1126/science.aal3099},
	doi = {10.1126/science.aal3099},
	language = {en},
	number = {6429},
	urldate = {2019-08-01},
	journal = {Science},
	author = {Wen, Xiao-Gang},
	month = feb,
	year = {2019},
	pages = {eaal3099},
}

@article{verresen_efficiently_2021,
	title = {Efficiently preparing {GHZ}, topological and fracton states by measuring cold atoms},
	url = {http://arxiv.org/abs/2112.03061},
	urldate = {2021-12-07},
	journal = {arXiv:2112.03061 [cond-mat, physics:quant-ph]},
	author = {Verresen, Ruben and Tantivasadakarn, Nathanan and Vishwanath, Ashvin},
	month = dec,
	year = {2021},
	note = {arXiv: 2112.03061},
}

@article{orus_tensor_2019,
	title = {Tensor networks for complex quantum systems},
	volume = {1},
	copyright = {2019 Springer Nature Limited},
	issn = {2522-5820},
	url = {https://www.nature.com/articles/s42254-019-0086-7},
	doi = {10.1038/s42254-019-0086-7},
	language = {en},
	number = {9},
	urldate = {2023-09-23},
	journal = {Nature Reviews Physics},
	author = {Orús, Román},
	month = sep,
	year = {2019},
	note = {Number: 9
Publisher: Nature Publishing Group},
	pages = {538--550},
}

@book{stanescu_introduction_2016,
	title = {Introduction to {Topological} {Quantum} {Matter} \& {Quantum} {Computation}},
	isbn = {978-1-4822-4594-3},
	language = {en},
	publisher = {CRC Press},
	author = {Stanescu, Tudor D.},
	month = dec,
	year = {2016},
	note = {Google-Books-ID: LZa\_DQAAQBAJ},
}

@article{kesselring_boundaries_2018,
	title = {The boundaries and twist defects of the color code and their applications to topological quantum computation},
	volume = {2},
	issn = {2521-327X},
	url = {http://arxiv.org/abs/1806.02820},
	doi = {10.22331/q-2018-10-19-101},
	language = {en},
	urldate = {2024-05-24},
	journal = {Quantum},
	author = {Kesselring, Markus S. and Pastawski, Fernando and Eisert, Jens and Brown, Benjamin J.},
	month = oct,
	year = {2018},
	note = {arXiv:1806.02820 [cond-mat, physics:quant-ph]},
	pages = {101},
}

@article{verresen_one-dimensional_2017,
	title = {One-dimensional symmetry protected topological phases and their transitions},
	volume = {96},
	copyright = {https://link.aps.org/licenses/aps-default-license},
	issn = {2469-9950, 2469-9969},
	url = {https://link.aps.org/doi/10.1103/PhysRevB.96.165124},
	doi = {10.1103/PhysRevB.96.165124},
	language = {en},
	number = {16},
	urldate = {2024-05-30},
	journal = {Physical Review B},
	author = {Verresen, Ruben and Moessner, Roderich and Pollmann, Frank},
	month = oct,
	year = {2017},
	pages = {165124},
}

@article{bombin_topological_2010,
	title = {Topological {Order} with a {Twist}: {Ising} {Anyons} from an {Abelian} {Model}},
	volume = {105},
	shorttitle = {Topological {Order} with a {Twist}},
	url = {https://link.aps.org/doi/10.1103/PhysRevLett.105.030403},
	doi = {10.1103/PhysRevLett.105.030403},
	number = {3},
	urldate = {2025-09-02},
	journal = {Physical Review Letters},
	author = {Bombin, H.},
	month = jul,
	year = {2010},
	note = {Publisher: American Physical Society},
	pages = {030403},
}

@article{yan_generalized_2024,
	title = {Generalized {Kitaev} spin liquid model and emergent twist defect},
	volume = {466},
	issn = {0003-4916},
	url = {https://www.sciencedirect.com/science/article/pii/S0003491624000903},
	doi = {10.1016/j.aop.2024.169682},
	urldate = {2024-06-12},
	journal = {Annals of Physics},
	author = {Yan, Bowen and Chen, Penghua and Cui, Shawn X.},
	month = jul,
	year = {2024},
	pages = {169682},
}

@article{you_projective_2012,
	title = {Projective non-{Abelian} statistics of dislocation defects in a {$ \mathbb{Z}_N$} rotor model},
	volume = {86},
	url = {https://link.aps.org/doi/10.1103/PhysRevB.86.161107},
	doi = {10.1103/PhysRevB.86.161107},
	number = {16},
	urldate = {2024-06-13},
	journal = {Physical Review B},
	author = {You, Yi-Zhuang and Wen, Xiao-Gang},
	month = oct,
	year = {2012},
	note = {Publisher: American Physical Society},
	pages = {161107},
}

@article{you_non-abelian_2019,
	title = {Non-{Abelian} defects in fracton phases of matter},
	volume = {100},
	url = {https://link.aps.org/doi/10.1103/PhysRevB.100.075148},
	doi = {10.1103/PhysRevB.100.075148},
	number = {7},
	urldate = {2024-09-25},
	journal = {Physical Review B},
	author = {You, Yizhi},
	month = aug,
	year = {2019},
	note = {Publisher: American Physical Society},
	pages = {075148},
}

@article{brown_topological_2013,
	title = {Topological {Entanglement} {Entropy} with a {Twist}},
	volume = {111},
	url = {https://link.aps.org/doi/10.1103/PhysRevLett.111.220402},
	doi = {10.1103/PhysRevLett.111.220402},
	number = {22},
	urldate = {2024-12-10},
	journal = {Physical Review Letters},
	author = {Brown, Benjamin J. and Bartlett, Stephen D. and Doherty, Andrew C. and Barrett, Sean D.},
	month = nov,
	year = {2013},
	note = {Publisher: American Physical Society},
	pages = {220402},
}

@article{chen_local_2010,
	title = {Local unitary transformation, long-range quantum entanglement, wave function renormalization, and topological order},
	volume = {82},
	copyright = {http://link.aps.org/licenses/aps-default-license},
	issn = {1098-0121, 1550-235X},
	url = {https://link.aps.org/doi/10.1103/PhysRevB.82.155138},
	doi = {10.1103/PhysRevB.82.155138},
	language = {en},
	number = {15},
	urldate = {2025-05-14},
	journal = {Physical Review B},
	author = {Chen, Xie and Gu, Zheng-Cheng and Wen, Xiao-Gang},
	month = oct,
	year = {2010},
	pages = {155138},
}

@article{chen_topological_2020,
	title = {Topological phase transition on the edge of two-dimensional {$\mathbb{Z}_2$} topological order},
	volume = {102},
	url = {https://link.aps.org/doi/10.1103/PhysRevB.102.045139},
	doi = {10.1103/PhysRevB.102.045139},
	number = {4},
	urldate = {2025-05-29},
	journal = {Physical Review B},
	author = {Chen, Wei-Qiang and Jian, Chao-Ming and Kong, Liang and You, Yi-Zhuang and Zheng, Hao},
	month = jul,
	year = {2020},
	note = {Publisher: American Physical Society},
	pages = {045139},
}

@article{garnerone_fidelity_2009,
	title = {Fidelity in topological quantum phases of matter},
	volume = {79},
	copyright = {http://link.aps.org/licenses/aps-default-license},
	issn = {1050-2947, 1094-1622},
	url = {https://link.aps.org/doi/10.1103/PhysRevA.79.032302},
	doi = {10.1103/PhysRevA.79.032302},
	language = {en},
	number = {3},
	urldate = {2025-08-19},
	journal = {Physical Review A},
	author = {Garnerone, Silvano and Abasto, Damian and Haas, Stephan and Zanardi, Paolo},
	month = mar,
	year = {2009},
	pages = {032302},
}

@article{barkeshli_symmetry_2019,
	title = {Symmetry fractionalization, defects, and gauging of topological phases},
	volume = {100},
	issn = {2469-9950, 2469-9969},
	url = {https://link.aps.org/doi/10.1103/PhysRevB.100.115147},
	doi = {10.1103/PhysRevB.100.115147},
	language = {en},
	number = {11},
	urldate = {2025-08-29},
	journal = {Physical Review B},
	author = {Barkeshli, Maissam and Bonderson, Parsa and Cheng, Meng and Wang, Zhenghan},
	month = sep,
	year = {2019},
	pages = {115147},
}

@article{terhal_quantum_2015,
	title = {Quantum error correction for quantum memories},
	volume = {87},
	issn = {0034-6861, 1539-0756},
	url = {https://link.aps.org/doi/10.1103/RevModPhys.87.307},
	doi = {10.1103/RevModPhys.87.307},
	language = {en},
	number = {2},
	urldate = {2020-02-25},
	journal = {Reviews of Modern Physics},
	author = {Terhal, Barbara M.},
	month = apr,
	year = {2015},
	pages = {307--346},
}

@article{zhou_quantum_2017,
	title = {Quantum spin liquid states},
	volume = {89},
	issn = {0034-6861, 1539-0756},
	url = {http://link.aps.org/doi/10.1103/RevModPhys.89.025003},
	doi = {10.1103/RevModPhys.89.025003},
	language = {en},
	number = {2},
	urldate = {2018-09-19},
	journal = {Reviews of Modern Physics},
	author = {Zhou, Yi and Kanoda, Kazushi and Ng, Tai-Kai},
	month = apr,
	year = {2017},
}

@article{kitaev_anyons_2006,
	title = {Anyons in an exactly solved model and beyond},
	volume = {321},
	issn = {00034916},
	url = {http://linkinghub.elsevier.com/retrieve/pii/S0003491605002381},
	doi = {10.1016/j.aop.2005.10.005},
	language = {en},
	number = {1},
	urldate = {2018-08-08},
	journal = {Annals of Physics},
	author = {Kitaev, Alexei},
	month = jan,
	year = {2006},
	pages = {2--111},
}

@article{fu_three_2018,
	title = {Three types of representation of spin in terms of {Majorana} fermions and an alternative solution of the {Kitaev} honeycomb model},
	volume = {97},
	url = {https://link.aps.org/doi/10.1103/PhysRevB.97.115142},
	doi = {10.1103/PhysRevB.97.115142},
	number = {11},
	urldate = {2025-04-30},
	journal = {Physical Review B},
	author = {Fu, Jianlong and Knolle, Johannes and Perkins, Natalia B.},
	month = mar,
	year = {2018},
	note = {Publisher: American Physical Society},
	pages = {115142},
}

@article{tang_unveiling_2021,
	title = {Unveiling quantum phase transitions by fidelity mapping},
	volume = {104},
	issn = {2469-9950, 2469-9969},
	url = {https://link.aps.org/doi/10.1103/PhysRevB.104.075142},
	doi = {10.1103/PhysRevB.104.075142},
	language = {en},
	number = {7},
	urldate = {2025-08-19},
	journal = {Physical Review B},
	author = {Tang, Ho-Kin and Marashli, Mohamad Ali and Yu, Wing Chi},
	month = aug,
	year = {2021},
	pages = {075142},
}

@article{you_fidelity_2007,
	title = {Fidelity, dynamic structure factor, and susceptibility in critical phenomena},
	volume = {76},
	copyright = {http://link.aps.org/licenses/aps-default-license},
	issn = {1539-3755, 1550-2376},
	url = {https://link.aps.org/doi/10.1103/PhysRevE.76.022101},
	doi = {10.1103/PhysRevE.76.022101},
	language = {en},
	number = {2},
	urldate = {2025-08-19},
	journal = {Physical Review E},
	author = {You, Wen-Long and Li, Ying-Wai and Gu, Shi-Jian},
	month = aug,
	year = {2007},
	pages = {022101},
}

@article{gu_fidelity_2010,
	title = {Fidelity approach to quantum phase transitions},
	volume = {24},
	issn = {0217-9792, 1793-6578},
	url = {http://arxiv.org/abs/0811.3127},
	doi = {10.1142/S0217979210056335},
	number = {23},
	urldate = {2025-08-19},
	journal = {International Journal of Modern Physics B},
	author = {Gu, Shi-Jian},
	month = sep,
	year = {2010},
	note = {arXiv:0811.3127 [quant-ph]},
	pages = {4371--4458},
}

@article{zanardi_information-theoretic_2007,
	title = {Information-{Theoretic} {Differential} {Geometry} of {Quantum} {Phase} {Transitions}},
	volume = {99},
	issn = {0031-9007, 1079-7114},
	url = {https://link.aps.org/doi/10.1103/PhysRevLett.99.100603},
	doi = {10.1103/PhysRevLett.99.100603},
	language = {en},
	number = {10},
	urldate = {2019-01-09},
	journal = {Physical Review Letters},
	author = {Zanardi, Paolo and Giorda, Paolo and Cozzini, Marco},
	month = sep,
	year = {2007},
}

@article{bridgeman_hand-waving_2017,
	title = {Hand-waving and interpretive dance: an introductory course on tensor networks},
	volume = {50},
	issn = {1751-8113, 1751-8121},
	shorttitle = {Hand-waving and interpretive dance},
	url = {http://stacks.iop.org/1751-8121/50/i=22/a=223001?key=crossref.20a5d8f2d9345e1b0655b8160d5d510b},
	doi = {10.1088/1751-8121/aa6dc3},
	language = {en},
	number = {22},
	urldate = {2019-05-14},
	journal = {Journal of Physics A: Mathematical and Theoretical},
	author = {Bridgeman, Jacob C and Chubb, Christopher T},
	month = jun,
	year = {2017},
	pages = {223001},
}

@article{lebreuilly_autonomous_2021,
	title = {Autonomous quantum error correction and quantum computation},
	url = {http://arxiv.org/abs/2103.05007},
	urldate = {2021-03-10},
	journal = {arXiv:2103.05007 [quant-ph]},
	author = {Lebreuilly, José and Noh, Kyungjoo and Wang, Chiao-Hsuan and Girvin, Steven M. and Jiang, Liang},
	month = mar,
	year = {2021},
	note = {arXiv: 2103.05007},
}

@article{bonilla_ataides_xzzx_2021,
	title = {The {XZZX} surface code},
	volume = {12},
	copyright = {2021 The Author(s)},
	issn = {2041-1723},
	url = {https://www.nature.com/articles/s41467-021-22274-1},
	doi = {10.1038/s41467-021-22274-1},
	language = {en},
	number = {1},
	urldate = {2026-03-03},
	journal = {Nature Communications},
	publisher = {Nature Publishing Group},
	author = {Bonilla Ataides, J. Pablo and Tuckett, David K. and Bartlett, Stephen D. and Flammia, Steven T. and Brown, Benjamin J.},
	month = apr,
	year = {2021},
	pages = {2172},
}

@article{schuch_matrix_2019,
	title = {Matrix product state algorithms for {Gaussian} fermionic states},
	volume = {100},
	url = {https://link.aps.org/doi/10.1103/PhysRevB.100.245121},
	doi = {10.1103/PhysRevB.100.245121},
	number = {24},
	urldate = {2026-03-03},
	journal = {Physical Review B},
	publisher = {American Physical Society},
	author = {Schuch, Norbert and Bauer, Bela},
	month = dec,
	year = {2019},
	pages = {245121},
}

\end{document}